\documentclass[prd,aps,twocolumn,floats,epsf]{revtex4}

\begin{document}

%\tighten
%\draft
%@twocolumnfalse\endcsname

\preprint{BROWN-HET-1392}

\title{Towards a New Proof of Anderson Localization}

\author{Robert Brandenberger$^{1)}$ and Walter Craig$^{2)}$}

\address{1) Department of Physics, McGill University, 3600 Rue Universit\'e, 
Montr\'eal H3A 2T8, Qu\'ebec, Canada\\
E-mail: rhb@physics.mcgill.ca}

\address{2) Department of Mathematics and Statistics, McMaster University \\
Hamilton, Ontario, L8S 4K1, Canada \\
E-mail: craig@math.mcmaster.ca}

\begin{abstract}
The wave function of a non-relativistic particle in a periodic potential
admits oscillatory solutions, the Bloch waves. In the presence of a 
random noise contribution to the potential the wave function is localized. 
We outline a new proof of this Anderson localization phenomenon in one spatial
dimension, extending the classical result to the case of a periodic background
potential. The proof makes use of techniques previously developed to study
the effects of noise on reheating in inflationary cosmology, 
employing methods of random matrix theory.
\end{abstract}

\maketitle
%\pacs{PACS numbers: 98.80.Cq, 98.70.Vc}
%\vspace*{1cm}
%]

\section{Introduction}

A classic problem in non-relativistic quantum mechanics is the
propagation of a particle, e.g. an electron, in a periodic potential,
set up e.g. by a lattice of ions. It is known that the time-independent 
Schr\"odinger equation for this problem admits modulated periodic 
solutions, the so-called Bloch waves \cite{Bloch}. 
In the real world, however, we expect that the potential will not
be perfectly periodic due to the presence of various types of disorder,
e.g. thermal noise. As was first shown by Anderson \cite{Anderson}, 
if the noise is modeled as a random contribution to the potential, then
for sufficiently large disorder the wave functions become localized. In
one spatial dimension, a stronger result holds: it can be shown
\cite{Mott} that for any amount of disorder, the wave functions
become exponentially localized, i.e. instead of being oscillatory, the
solutions now are exponentially decaying in space about the location
of the center of the wave packet. This result was extended to
two spatial dimensions in \cite{Abrahams}. Anderson localization
is reviewed in \cite{Rev1,Rev2,Rev3}. A mathematically rigorous
proof of one dimensional Anderson localization for constant background
potential is given in \cite{GMP}, and there is an extensive mathematical
literature on the subject.

As is well known, in one spatial dimension the time-independent
Schr\"odinger equation can be mapped into a classical time-dependent
wave equation by a simple variable transformation in which the
wave function $\Psi$ becomes a classical field $\chi$ and the 
spatial variable $x$ is transformed to time $t$. After this
mapping, the Schr\"odinger equation for the wave function $\Psi$
in a potential which is the sum of a periodic term $V_p(x)$ and a random
noise term $V_R(x)$ becomes the relativistic wave equation (in momentum 
space) for a scalar field with a mass which contains one contribution
to the mass which is periodically varying in time, and a second contribution 
which is random in time. 

The inflationary scenario \cite{Guth} (see also \cite{RHBrev}
for a recent review emphasizing both successes and problems of
inflation), the current paradigm of early
universe cosmology, is based on the dynamics of classical scalar fields
coupled as matter source to Einstein's theory of General Relativity.
According to the inflationary universe scenario, there is a period in
the very early universe in which space expands almost exponentially.
This is obtained by making use of a scalar field $\varphi$ whose 
energy functional is dominated by a potential energy density
contribution which is almost constant in time. Regular matter can
be modeled as a second scalar field $\chi$. 

During the period of inflation, an exponentially increasing fraction of the
energy density of the universe is stored in the field $\varphi$. Hence,
to make contact with what is observed today, a phase at the end
of the period of inflation when the energy transfers from $\varphi$
to regular matter is crucial. This phase is called the ``reheating
stage'', and works in the following way. The period of inflation
terminates once $\varphi$ begins to oscillate about the minimum of
its potential. Due to a resonant coupling between $\varphi$ and the matter
field $\chi$, energy can be transferred from $\varphi$ to $\chi$. 
The theory of reheating is obtained by studying the evolution of
$\chi$ in the presence of an oscillating $\varphi$ field.

The equation of motion for $\chi$ is the Klein-Gordon equation with
a correction term due to the coupling with $\varphi$.
Treating $\chi$ at the linearized level, each Fourier mode
of $\chi$ evolved independently. Neglecting
the effects of the expansion of the universe \footnote{This is self-consistent
if the period of energy transfer turns out to be short compared to the
time scale of the expansion of space.}, and for the specific
coupling between the two fields which we specify below, the equation of
motion for such a Fourier mode of $\chi$ is that of a harmonic
oscillator whose mass has two time-dependent contributions, one periodic 
in time (due to the oscillatory dynamics of $\varphi$ during the
reheating period, the other being an aperiodic random noise term (due e.g.
to quantum fluctuations in $\varphi$). 

In the absence of noise, the equation of motion is the Mathieu equation
and falls into the class of equations first studied by Hill, Floquet and
Legendre \cite{Hill}. As a function of the value of the bare mass (the term in
the mass independent of time), there are stability bands (for which
the solutions are oscillatory) and instability bands (for which
the solutions are characterized by overall exponential growth or
decay. The coefficient describing the exponential growth is called
the ``Lyapunov exponent''. Applied to the case of cosmology, then in
the absence of noise there are bands of Fourier modes for which the
mass lies in the instability band and for which there is resonant
increase in the amplitude of $\chi$, which in terms of physics corresponds
to resonant production of particles \cite{TB1}.

Not too long ago \cite{Craig1}, the  effects of a particular type 
of random noise on the resonant production of particles during
inflationary reheating was studied.
It was shown that the Lyapunov exponent $\mu_k(q)$ which
describes the exponential growth of each Fourier mode $k$ of the
scalar field in the presence of the noise $q$ is strictly larger than the 
corresponding Lyapunov exponent $\mu_k(0)$ in the absence of noise 
\footnote{Non-perturbatively, this was only shown for values of $k$ in the
stability bands. Perturbatively, the statement also holds for values of
$k$ in the instability bands \cite{Craig4}. What is relevant for this
paper is the result for values of $k$ in the stability bands.}.
In particular, this implies that in the presence of noise, every
Fourier mode grows exponentially. The stability bands which are present
in the absence of noise disappear. 

In this Letter, we review the above analysis, translate the results
into the language of the time-independent Schr\"odinger equation,
and in this way immediately obtain a new proof of Anderson localization
which extends to the case of a periodic background potential. 

\vskip.5cm
\section{Review of Results on the Effects of Noise on Reheating in
Inflationary Cosmology}

We begin by reviewing the results of \cite{Craig1}. We consider
the simplest model which describes both the inflationary phase
of the very early Universe and the period of ``reheating'' which
terminates the phase of inflation and during which the energy of
the Universe is transferred to regular matter. This model contains
two scalar matter fields, $\varphi$ and $\chi$. The first,
$\varphi$, has a large
potential energy which is changing only very slowly in time,
dominates the total energy density of the Universe and thus gives
rise to inflation \cite{Guth}. The second scalar field, $\chi$,
represents regular matter. It is in its vacuum state initially and
gets excited by a coupling with $\varphi$ during the final
stages of inflation when $\varphi(t)$ oscillates about the minimum
of its potential \footnote{This toy model was already used in the
first studies of reheating in inflationary cosmology \cite{Dolgov,AFW}.}.
It is of great interest in cosmology to study the dynamics of matter
production at the end of inflation.

We will assume that the interaction Lagrangian
density takes the form \footnote{The quantity $g$ is a constant which
has dimensions of mass. It is the coupling constant.}
\begin{equation}
{\cal L}_{\rm int} \, = \, {1 \over 2} g \varphi \chi^2 \, ,
\end{equation}
and that the Lagrangian for $\chi$ alone is that of a free scalar field
with mass $m_{\chi}$. We will also consider $\varphi$ to be
spatially homogeneous. In the absence of quantum fluctuations, this
is a reasonable assumption since during the period of inflation, spatial
fluctuations are red-shifted exponentially. In this case, the equation of 
motion for $\chi$ can be solved independently for each Fourier mode.
In the absence of expansion of the Universe, the resulting equation
is \footnote{The expansion of the Universe can be taken into account
\cite{KLS1,TB2,KLS2} without affecting the result that there are
exponential instabilities.}  
\begin{equation}
{\ddot \chi_k} + \left[\omega_k^2 + g \varphi(t) \right] \chi_k \, = 0
\, , \label{KG1}
\end{equation}
where $\omega_k^2 = k^2 +m_{\chi}^2$. 

At the end of the period of inflation, the scalar field $\varphi$ is
oscillating about its ground state value, which we without loss of
generality can take to be $\varphi = 0$. The frequency $\omega$
of the oscillation is set by the mass of $\varphi$. In this case,
the equation (\ref{KG1}) has the form of a Mathieu equation. As is well
known (see e.g. \cite{Hill,LL1,Arnold}), there are bands of values 
of $\omega_k$ for which $\chi_k$ grows exponentially. The exponential 
growth is governed by a Lyapunov exponent $\mu(k)$, which can be
extracted from the time evolution of $\chi_k(t)$ as follows:
\begin{equation} \label{Floquet1}
\mu_k \, = \, {\rm lim}_{t \rightarrow \infty} {1 \over t} log |\chi_k(t)|
\, .
\end{equation}

The time scale of
exponential growth is in many concrete inflationary models much
shorter than the typical expansion time of the Universe, thus
justifying the neglect of the expansion. Thus \cite{TB1}, the parametric
instability leads to a very efficient energy transfer \footnote{At
some point, the back-reaction of the energy in the $\chi$ field
on the dynamics of space-time will become important, and this could
terminate the energy transfer \cite{Zibin}.}

However, in cosmology we expect fluctuations of thermal or quantum
nature to be super-imposed on the homogeneous oscillation of $\varphi$.
In fact, we believe that quantum vacuum fluctuations of $\varphi$ in
inflationary cosmology are the seeds for the inhomogeneities in the
galaxy distribution and the anisotropies in the temperature of the cosmic
microwave background observed today. Thus, it becomes important to study
the sensitivity of the parametric resonance instability of (\ref{KG1})
in the presence of an oscillating $\varphi(t)$ to the presence of
random noise in $\varphi(t)$. In \cite{Craig1}, this problem was
studied for homogeneous, aperiodic noise, modeled as the addition
of a stochastic contribution to the time dependent mass. More
precisely, we studied the equation
\begin{equation}
{\ddot \chi_k} + \left[\omega_k^2 + p(\omega t) + q(t) \right] \chi_k \, = 0
\, , \label{KG2}
\end{equation}
where $p$ is a periodic function with period $2 \pi$ and $q(t)$ is
an aperiodic, random noise contribution \footnote{The mathematically much
more complicated problem of inhomogeneous noise, in which case the
inhomogeneity in the noise couples the different Fourier modes of
$\chi$, thus leading to a problem in the field of partial differential
equations, was studied in \cite{Craig2}.}

To derive rigorous inequalities on the magnitudes of the Lyapunov exponent
with and without noise, it is necessary to make certain assumptions
on the noise $q(t)$. We consider 
noise drawn from some sample space $\Omega = C({\cal R})$ with
a translationally invariant probability measure $dP(\kappa)$, where
$\kappa$ labels an element in $\Omega$, and assume:
\begin{itemize}
\item{} The noise is ergodic, i.e. the ensemble average of a function of the 
noise equals the time average for almost all realizations of the noise.
\item{} The noise is uncorrelated in time on scales larger than $T$,
the period of the oscillatory contribution to the mass in (\ref{KG2}).
\item{} Restricting the noise to the time interval $0 \leq t < T$,
the samples ${q(t; \kappa): 0 \leq t < T}$ within the support of
the probability measure fill a neighborhood of the origin in $\Omega$.
\end{itemize}
Given these assumptions, we were able to use a theorem of random matrix
theory (applied to the transfer matrix corresponding to the differential
equation (\ref{KG2})) to prove that for almost all realizations of the
noise
\begin{equation} \label{result}
\mu(q) > \mu(0) 
\end{equation}
for all values of the momentum $k$. In particular, this implies that
the stability bands of the Mathieu equation disappear once noise
of the type considered here is added to the system. The noise
in fact strengthens the instability. Note that the result (\ref{result})
was proven non-perturbatively in \cite{Craig1} for values of $k$ in the
stability bands, the case of interest to us here. We conjecture that
the result is true in general, and have been able to show this at least
perturbatively \cite{Craig4}.

\vskip.5cm
\section{Application to Anderson Localization}

Let us start from Eq. (\ref{KG2}) and transform variables. We
replace the time coordinate $t$ by a spatial coordinate $x$
(space being infinite, i.e. ${\cal R}$), 
and substitute the field variable $\chi$ by the variable $\psi$
representing a wave function. With these substitutions, Eq. (\ref{KG2})
becomes the time-independent Schr\"odinger equation for the wave
function $\psi$ in one spatial dimension for a system with a
potential which is the superposition of a periodic piece $V_p(x)$
(coming from $p(\omega t)$ in (\ref{KG2}) and a random noise
piece $V_R(x)$ coming from $q(t)$ in (\ref{KG2}):
\begin{equation} \label{Seq}
H \psi \, = \, E \psi
\end{equation}
with
\begin{equation} \label{Ham}
H \, = \, - {1 \over {2 m}}{{\partial^2} \over {{\partial x}^2}} + V_p(x)
+ V_R(x) \, .
\end{equation}
The energy eigenvalue $E$ is given by the momentum $k$ of (\ref{KG2}). We
use units in which $\hbar = 1$.

In the absence of noise, the spectrum of the Schr\"odinger equation
(\ref{Seq}) consists of bands of continuous spectrum, for which bounded
quasi-periodic (modulated periodic) solutions exist, alternating with
instability gaps (resonance intervals) in which the solution behavior
is exponential. In both regions the solutions, the so-called Bloch waves
\cite{Bloch}, exhibit Floquet behavior, namely they are of the form 
$\psi(x) = \exp(\pm(\mu_E + i \alpha_E)x)p(x)$, where $p(x)$ is
periodic of the same period as the potential $V_p$. The Floquet
exponent $m_E =  (\mu_E + i \alpha_E)$ has real part the Lyapunov
exponent and imaginary part the rotation number of the solution
$\psi(x)$. The Lyapunov exponent vanishes in the stable bands, while
the rotation number takes (a fixed multiple of) integer values in the
gaps. In the classical field theory problem these bands of spectrum
correspond precisely to the stability bands of the Mathieu equation
(\ref{KG2}) (for $q(t) = 0$).

%%%%%%%%%%%%
%% In the absence of noise, the Schr\"odinger equation (\ref{Seq}) admits
%% energy intervals for which periodic solutions, the so-called Bloch waves
%% \cite{Bloch}, exist. In the classical field theory problem these energy
%% intervals correspond to the stability bands of the Mathieu equation
%% (\ref{KG2}) (for $q(t) = 0$).
%%%%%%%%%%

Let us now turn on noise satisfying the conditions listed in the
previous section and look for solutions of (\ref{Seq}). Due to
the translation invariance of the problem, we can take $x = 0$ to
be the point at which $\psi$ takes on its maximum. In the language
of the quantum mechanics problem, the Lyapunov exponent can be 
extracted as follows:
\begin{equation} \label{Floquet2}
\mu_E \, = - \, {\rm lim}_{x \rightarrow \infty} {1 \over x} log |\psi(x)|
\, .
\end{equation}
Since the wave function in quantum mechanics must be normalized, no
exponentially growing solutions are allowed. An eigenstate corresponds
to a wave function which does not grow either for $x \rightarrow \infty$
or $x \rightarrow - \infty$. Thus, a positive real Lyapunov
exponent corresponds to an exponentially decaying wave function,
whereas an imaginary Floquet exponent with absolute value $1$ corresponds
to Bloch waves. A standard mathematical argument \cite{GMP,CKM,FMSS,CL} 
shows that the eigenstates are dense and the spectrum is pure point.

If we consider an energy eigenvalue $E$ which lies in a conduction band
(i.e. an energy band for which Bloch wave solutions exist) in the
absence of noise, then our result (\ref{result}) implies that as soon
as a random noise term is added to the potential in (\ref{Ham}),
the Lyapunov exponent becomes positive, and that thus the
corresponding wave function are exponentially decaying. This means
that the addition of any noise satisfying the conditions listed in the
previous section exponentially localizes the wave function. The 
localization length is inversely proportional to the Lyapunov exponent.  

\vskip.5cm
\section{Discussion}

We have shown that the time-independent Schr\"odinger
equation for the non-relativistic Hamiltonian (\ref{Ham}) does not
admit Bloch wave solutions for any amplitude of the noise. Given
a wave function centered at some point $x$ (which we can without loss
of generality take to be $x = 0$), we have shown that the wave function
decays exponentially away from $x = 0$. Thus, we have given a new
proof of the classic result \cite{Mott} of exponential localization of
states in one spatial dimension which extends the known result to the
case of a periodic background potential.

Our methods are based on translating the time-independent quantum mechanical
problem to a classical dynamical systems problem, and thus are
only applicable to the case of one spatial dimension and cannot be used to
study Anderson localization in a higher number of spatial dimensions. 

The dynamical systems problem turns out to be a problem concerning the effects
of noise on the resonant production of particles during the reheating
phase of inflationary cosmology. In mathematical language, our problem
concerns adding aperiodic noise to the mass term in the Mathieu
equation. We have shown that the Lyapunov exponent increases for any
amplitude of the noise. Thus, the solutions exhibit exponential
behavior both forwards and backwards in time (starting from the 
initial time $t = 0$ corresponding to $x = 0$ for the quantum problem). 
Normalizability of the wave function in the quantum problem implies
that the eigenstates are precisely those solutions which decay away 
from the source point.

\vskip.5cm
\centerline{\bf Acknowledgments}
\vskip.5cm
We are grateful to Profs. A. Maia and V. Zanchin for collaborating
with us on the earlier work on the effects of noise in inflationary
cosmology. One of us (RB) wishes to thank Prof. T. D. Lee for
asking key questions during a seminar at Columbia University. 
We would like to acknowledge the hospitality of the
Fields Institute for Research in Mathematical Sciences during
its 2003-2004 Thematic Program on Partial Differential Equations
which brought us together to study this problem.
This research is partly supported by NSERC Discovery Grants to R.B.
and W.C., and by the Canada Research Chair program.

\end{document}